# Topological phonons in an inhomogeneously strained silicon-2: Evidence of spin-momentum locking

Anand Katailiha[1‡], Paul C. Lou[1‡], Ravindra G. Bhardwaj[1‡], Ward P. Beyermann[2], and Sandeep Kumar[1,*]

[1]Department of Mechanical engineering, University of California, Riverside, CA 92521, USA

[2] Department of Physics and Astronomy, University of California, Riverside, CA 92521, USA

[*]Corresponding author

Email: sandeep.suk191@gmail.com

**Abstract**

In this study, we report first experimental evidence of spin-momentum locking in the topological phonons in an inhomogeneously strained Si thin film. The spin-momentum locking in the topological phonons lead to a longitudinal spin texture or spatially inhomogeneous spin distribution in the freestanding sample structure. The spin texture was uncovered using location dependent Hall effect and planar Hall effect measurement. The charge carrier density and anomalous Hall resistance showed a linear behavior along the length of the sample. Similarly, the planar Hall resistance related with the spin dependent scattering was also found to be different at two different location along the length of the sample. The spin-momentum locking also gave rise to transverse thermal spin current and spin-Nernst magneto thermopower response, which was uncovered using angle dependent longitudinal second harmonic measurement. The magneto thermopower response was also a function of crystallography of the Si sample where the sign of the response was opposite for <110> and <100> aligned samples. The spin-momentum locking in topological phonons may give rise to large spin dependent response at and above room temperature, which can pave the way for energy efficient spintronics and spin-caloritronics devices.

The superposition of polarization and dynamical polarization of optical phonons give rise to dynamical multiferroicity[1-3], which can be described by following equation:

$$\boldsymbol{M}_t \propto \boldsymbol{P} \times \partial_t \boldsymbol{P} \tag{1}$$

where $\boldsymbol{M}_t$ and $\boldsymbol{P}$ are temporal magnetic moment and polarization of optical phonons, respectively. Recently, dynamical multiferroicity was reported in highly doped (conducting) Si thin films under an applied strain gradient[4]. which can be described as:

$$\boldsymbol{M}_t \propto \boldsymbol{P}_{FE} \times \partial_t \boldsymbol{P} \tag{2}$$

where $\boldsymbol{P}_{FE}$ is the flexoelectronic effect. However, the strain gradient also creates an inhomogeneous medium where the phonon dispersion and frequency are now a function of spatial coordinates and, as a consequence, topological phonons arise in the Si thin films. In the previous study (part 1), we reported experimental evidence of topological phonons in an inhomogeneously strained Si thin films. We also reported transverse spin Nernst effect (TSNE) behavior where the sign of the response changed along the length of the sample. That sign change can be attributed to the spin-momentum locking of topological phonons. However, the mechanistic origin and direct experimental evidence was lacking, which motivated us to undertake this work.

For the topological phonons in an inhomogeneous medium, the time evolution of the phonon polarization is a function of Berry gauge potential ($\boldsymbol{A}$) and momentum ($\boldsymbol{p}$) as described by the following equation:

$$\partial_t \boldsymbol{P} \propto f(\boldsymbol{A}, \boldsymbol{p}) \tag{3}$$

As a consequence, the temporal magnetic moment (or spin angular momentum) can be described as:

$$M_t \propto P_{FE} \times f(A, p) \qquad (4)$$

From this relationship, we deduced that the topological phonons must have spin-momentum locking similar to the electronic spin-momentum locking in the topological insulator surface states. Our contention was also supported by theoretical prediction by Long et al.[5] The topological phonons having opposite helicity will carry opposite spin angular moment (or temporal magnetic moment) due to spin-momentum locking. As a consequence, the counterpropagating (opposite helicity) topological phonons must give rise to a spin texture along the length, width and thickness of a 3-dimensional thin film sample as shown in Figure 1 (a). Hence, the magnetic field dependent transport measurements as a function of spatial coordinates can be used to uncover the spin texture. In this study, we report experimental evidence of spin-momentum locking in topological phonons. The spin-momentum locking of topological phonons was uncovered using location dependent measurement of Hall effect and planar Hall effect, which showed position dependent charge carrier concentration, anomalous Hall resistance and planar Hall coefficient. Additionally, spin-momentum locking of topological phonons also gave rise to longitudinal spin-Nernst effect response, which was measured using angle dependent longitudinal second harmonic voltage.

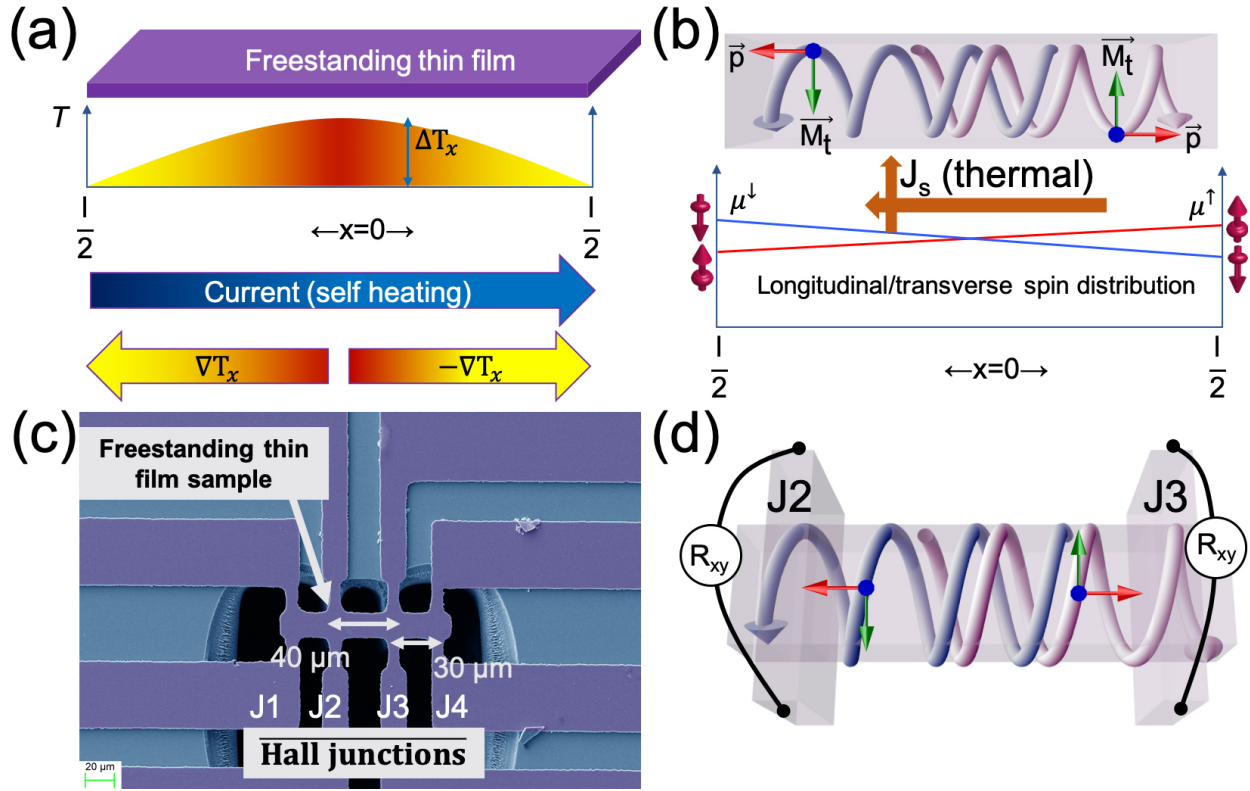

Figure 1. (a) A schematic showing the expected longitudinal temperature distribution in a freestanding sample where heat flows from the center to edges, (b) a schematic showing the topological phonons having spin-momentum locking that gives rise to longitudinal spin distribution, (c) a representative scanning electron micrograph showing the freestanding thin film structure having four Hall junctions, and (d) a schematic showing the experimental scheme where transverse resistance behavior was measured at two inner junctions J2 and J3.

In the freestanding (or suspended) sample, a longitudinal charge current leads to a parabolic (approximately) temperature distribution along the length of the sample as shown in Figure 1 (b) because of Joule heating. As a consequence, the heat flow will occur towards the boundaries from the center of the sample as shown in Figure 1(b). This, in turn, will give rise to the required spin texture due to the spin momentum locking of left and right propagating topological phonons. A spatial measurement can be used to measure the resulting effects. We tested our hypothesis using a freestanding Pd (1 nm)/Ni$_{80}$Fe$_{20}$ (Py)(25 nm)/MgO (1.8 nm)/Native oxide/p-Si (2 µm) thin film sample as shown in Figure 1 (c). The device structure has four Hall junctions that allow us to measure the longitudinal as well as transverse responses. Based on our hypothesis, the transverse resistance measured at J2 and J3, as shown in Figure 1 (d), should have spatially inhomogeneous spin dependent behavior due to spin-momentum locking.

First, we measured the Hall resistance response at J2 and J3 as a function of applied current (0.5 mA, 5 mA and 10 mA) for an applied magnetic field sweep from 3 T to -3 T as shown in Figure 2 (a). The applied current changed the Joule heating as well as strain gradient due to thermal expansion mismatch in the layered sample, which allowed us to uncover the relationship between inhomogeneity and spin-momentum locking of the topological phonons. The overall transverse responses reduced as a function of applied current, which was attributed to the strain gradient mediated flexoelectronic effect. Using linear fit, we estimated the Hall resistances for positive and negative magnetic fields and consequently the charge carrier concentrations were also estimated. Usually, the Hall resistance (slope) is same for both positive and negative magnetic fields. However, it is not true for Py thin films where slope for positive and negative magnetic fields are different, which can be attributed to the spin dependent skew scattering.

In our sample, we interpreted the calculated charge carrier concentration for positive and negative magnetic fields as corresponding to the spin-up and the spin-down charge carriers, respectively, as shown in Figure 2 (a). Hence, the charge carrier concentration should be significantly different when measurements are carried out at two different locations along the length due to spin-momentum locking from topological phonons in Si layer as shown in Figure 1 (a). The calculated values of the charge carrier concentrations and the anomalous Hall resistance are listed in the Table 1. We plotted the charge carrier concentration and anomalous Hall resistance measured at J2 and J3 to understand the longitudinal behavior as shown in Figure 2 (b). At 0.5 mA of applied current, the charge carrier concentration was larger at J2 as compared to J3 for both positive (spin-up) and negative (spin-down) magnetic fields as shown in Figure 2 (b) and Table 1. For 5 mA and 10 mA of applied current, the charge carrier concentration behavior was reversed and the values at J3 were larger than that at J2 as shown in Figure 2 (b). This was the first experimental evidence of the spatially inhomogeneous spin distribution due to spin-momentum locking as hypothesized.

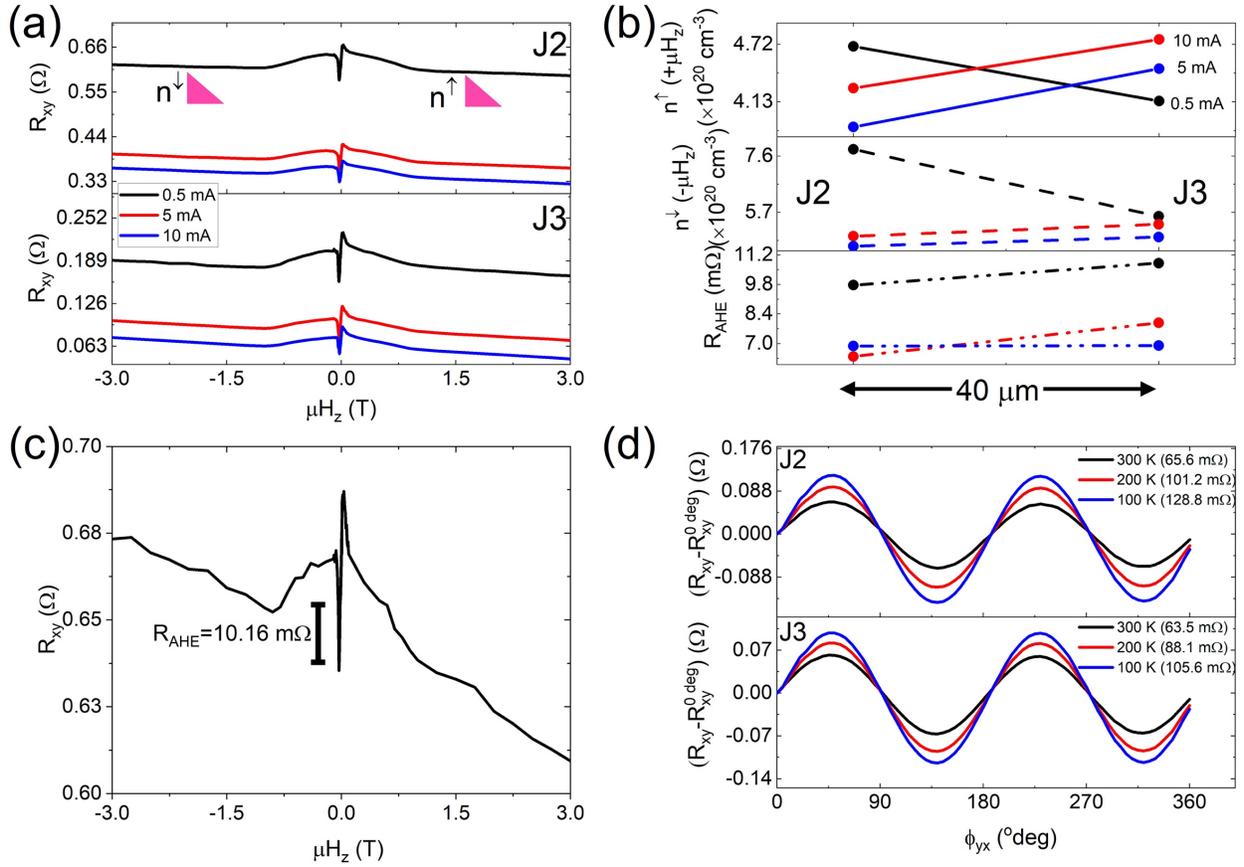

Figure 2. (a) the Hall effect measurement as a function of magnetic field from 3 T to -3 T at junctions J2 and J3 in Py (25 nm) MgO/p-Si (2 µm) sample. The measurement was carried out as a function of current at 0.5 mA, 5 mA and 10 mA to measure the effect of increased strain gradient from self-heating. (b) The longitudinal behavior of the charge carrier concentration of spin-up and spin down charge carriers (electrons) and anomalous Hall resistance. (c) The Hall resistance measured at 375 K from 3 T to -3 T magnetic field. (d) The planar Hall effect response measured using angle dependent transverse resistance at J2 and J3 as a function of temperature at 300 K, 200 K and 100 K.

Similarly, the anomalous Hall resistance was larger at J3 as compared to J2 for 0.5 mA of current as shown in Figure 2 (b) and Table 1. The anomalous Hall resistance reduced with increased applied current. At 10 mA, the anomalous Hall resistances were similar at both locations (J2 and J3). This evolution of the anomalous Hall resistance was attributed to the spatially inhomogeneous spin distribution from spin-momentum locking leading to spatially inhomogeneous magnetic moment. We estimated a temperature rise of ~70 K due to 10 mA of current. We, then, measured the transverse resistance response at J2 at 375 K for an applied current of 0.5 mA. The anomalous Hall resistance was estimated to be 10.16 mΩ as shown in Figure 2 (c), which was significantly larger than 6.88 mΩ measured at 10 mA and 300 K. This measurement showed that heating was not the underlying cause of the change in charge carrier concentration and anomalous Hall resistance as a function of applied current measured at 300 K and shown in Figure 2 (a,b).

Table 1. The anomalous Hall resistance and carrier concentration estimated for positive and negative magnetic fields at junctions J2 and J3.

|  | Measurement | J2 | | J3 | |
|---|---|---|---|---|---|
|  |  | Positive $\mu H_z$ | Negative $\mu H_z$ | Positive $\mu H_z$ | Negative $\mu H_z$ |
| 0.5 mA | $R_{AHE}$ (mΩ) | 9.77 | | 10.81 | |
|  | Carrier concentration (cm$^{-3}$) | 4.7×10$^{20}$ | 7.83×10$^{20}$ | 4.137×10$^{20}$ | 5.56×10$^{20}$ |
| 5 mA | $R_{AHE}$ (mΩ) | 6.38 | | 7.98 | |
|  | Carrier concentration (cm$^{-3}$) | 4.27×10$^{20}$ | 4.89×10$^{20}$ | 4.77×10$^{20}$ | 5.29×10$^{20}$ |
|  | $R_{AHE}$ (mΩ) | 6.88 | | 6.9 | |

| 10 mA | Carrier concentration (cm$^{-3}$) | 3.87×10$^{20}$ | 4.55×10$^{20}$ | 4.47×10$^{20}$ | 4.86×10$^{20}$ |

We, then, measured the angle dependent transverse resistance in the plane of the film at J2 and J3 as a function of temperature from 300 K to 100 K to further support our argument of longitudinal spin texture. The measured responses demonstrated the planar Hall effect (PHE) resistance as shown in Figure 2 (d). Similar to the charge carrier concentration, the PHE resistances were larger at J2 as compared to the values at J3 as shown in Figure 2 (d). The PHE response arises from the resistivity anisotropy from spin dependent scattering due to an in-plane magnetic field. The PHE behavior in Py thin films is well characterized and is not expected to be the function of location. Whereas our measurements clearly demonstrated different PHE responses at two different locations. In addition, the difference in responses increased to ~10% as the temperature was reduced to 100 K as shown in Figure 2 (d). The PHE measurement, anomalous Hall and Hall effect measurements demonstrated spatially inhomogeneous spin dependent behavior. This can also be called as inhomogeneous magnetoelectronic effect due to dynamical multiferroicity as presented in equation 4, which was the first evidence of spin-momentum locking.

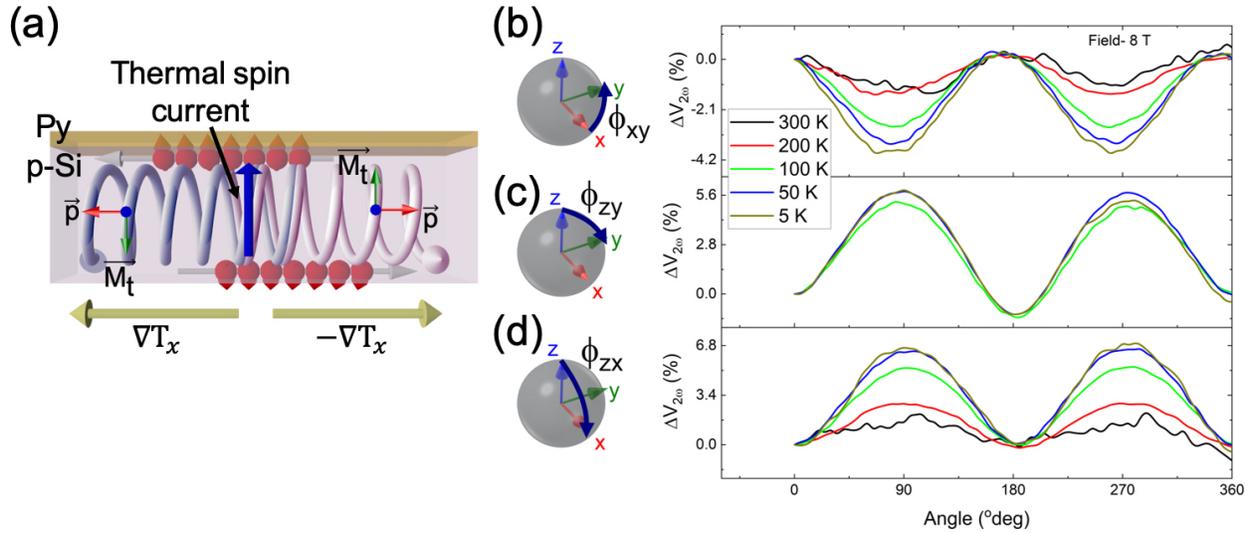

Figure 3. (a) Schematic showing the origin of transverse thermal spin current in self-heated freestanding sample due to spin-momentum locking of topological phonons. The angle dependent longitudinal second harmonic response in Py/MgO/p-Si sample measured at 8 T magnetic field as a function of temperature (b) response in xy- plane, (c) zy- plane, and (d) zx- plane.

As stated earlier, longitudinal charge current also leads to longitudinal heat transport from center of the structure to boundaries in the freestanding thin film structure. In the normal freestanding samples, the symmetry of the temperature gradient leads to absence of thermopower response in the second harmonic measurement. However, the topological phonons having opposite helicity will shift in opposite direction giving rise to topological spin-Hall effect of phonons[6] irrespective of the direction of heat flow. Our argument was also supported by the magnetoresistance behavior presented in part 1 where topological edge states like spin accumulation was reported. Due to the spin dependent electron-phonon interactions, topological phonon transport will also lead to longitudinal and transverse thermoelectric spin current due to spin-momentum locking as shown in Figure 3 (a), which is also known as spin-Nernst effect[7-9] a thermal counterpart of spin-Hall effect. The spin-Hall effect behavior is measured using spin-Hall magnetoresistance (SMR)[10] in bilayer consisting of ferromagnetic layer and heavy metal. It arises due to modulation in the resistance due to relative orientation between ferromagnetic magnetization and spin polarization of the transverse spin current due to spin-Hall effect. In addition, the unidirectional SMR[11] response also arises due to spin accumulation at the interface, which is observed in the longitudinal second harmonic measurement. In case of topological phonons, the SMR behavior arises due to spin dependent phonon skew scattering in the charge transport, which has been demonstrated recently[12]. In this study, the spin Nernst effect was expected to be observed in the angle dependent longitudinal second harmonic response. To prove our hypothesis, the angle dependent longitudinal $V_{2\omega}$ response was measured in xy–, zx– and zy–planes as shown in Figure 3 (b-d) similar to the traditional SMR measurement scheme. The measurement was carried out at 300 K, 200 K, 100 K, 50 K and 5 K and an applied magnetic field of 8 T as shown in Figure 3 (b-d). The angle dependent response in zy-plane at 300 K and 200 K were analyzed separately and are not plotted in Figure 3 (c). The measured longitudinal $V_{2\omega}$ responses showed $\sin^2 \phi$ symmetry expected for spin Nernst magneto-thermopower (SNMTP)[9] responses. Such behavior has not been reported either in the Py or the Si thin films because there was no longitudinal temperature difference. This response can only arise from topological phonons having spin-momentum locking.

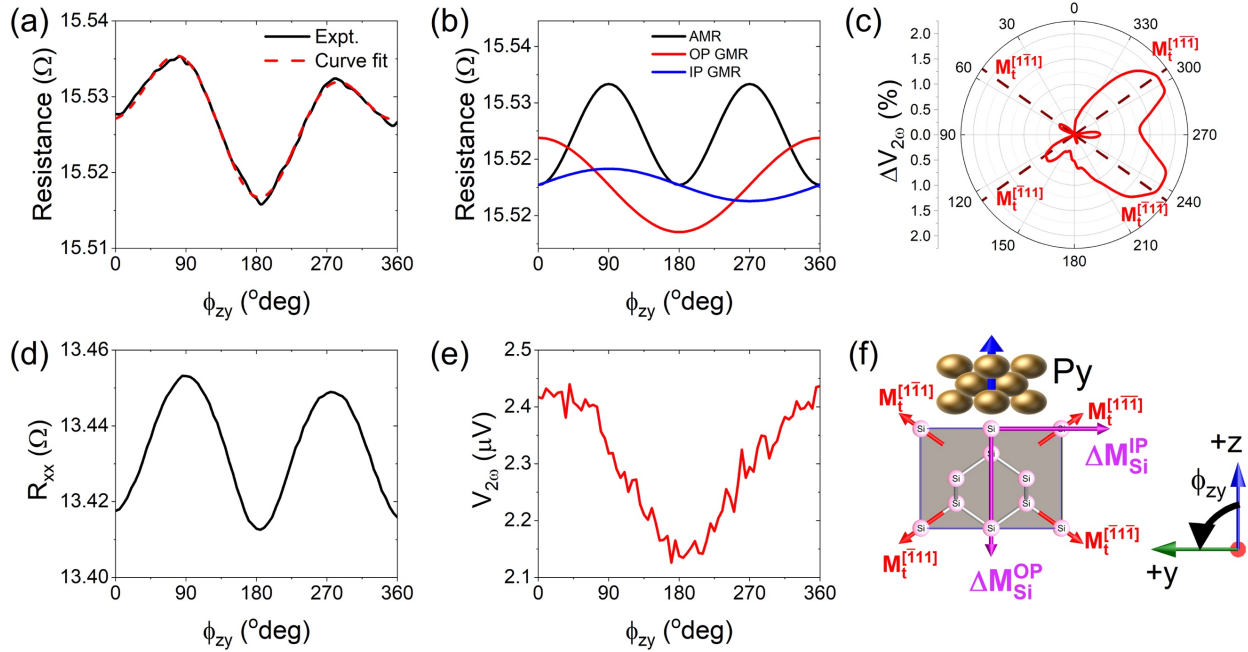

Figure 4. (a) the angle-dependent longitudinal resistance for an applied magnetic field of 8 T at 300 K. (b) The simulated responses due to AMR and GMR in CIP configuration that gave rise to the total response in (a). (c) The angle-dependent $V_{2\omega}$ (non-reciprocal) response at 8 T and 300 K showing the temporal magnetic moment due to dynamical multiferroicity. (d) the angle-dependent longitudinal resistance for an applied magnetic field of 8 T at 300 K. (e) the angle-dependent longitudinal second harmonic for an applied magnetic field of 8 T at 200 K. (f) And a schematic showing the temporal magnetic moments that gave rise to net magnetic moment in the -y and -z-directions and also responsible for the GMR responses.

We, then, analyzed the angle dependent resistance and longitudinal $V_{2\omega}$ responses in zy-plane measured at 300 K and 200 K for an applied magnetic field of 8 T. The angle dependent magnetoresistance response measured at 300 K is shown in Figure 4 (a). The measured resistance behavior was simulated using anisotropic magnetoresistance (AMR) with $\sin^2\theta$ behavior and two sinusoidal responses (sine and cosine each), as shown in Figure 4 (b). The AMR response was attributed to the ferromagnetic (Py) layer. The two sinusoidal responses were attributed to the giant magnetoresistance (GMR) in the current-in-plane (CIP) geometry. The two responses required two net magnetic moments in the p-Si layer: one along the -y direction (GMR-0.067%) and the other in the -z direction (GMR-0.14%). The angle-dependent longitudinal second harmonic response was associated with the dynamical multiferroicity and temporal magnetic moment in our case[13]. The measured angle-dependent $V_{2\omega}$ response exhibited behavior having four maximums along the <111> directions, as shown in Figure 4 (c). The magnitudes of the temporal magnetic moments were different in different symmetry directions, and in our sample, $M_t^{[\bar{1}11]} < M_t^{[1\bar{1}\bar{1}]}$ and $M_t^{[1\bar{1}11]} < M_t^{[\bar{1}1\bar{1}]}$, as shown in Figure 4 (c). Thus, the Si layer had a magnetic moment component in the -y direction and in the -z direction, as deduced from the GMR response and as shown in Figure 4 (d). The dynamical multiferroicity disappeared when the temperature was reduced 200 K as shown in Figure 4 (e,f). The GMR responses were also significantly smaller as shown in Figure 4 (e) and disappeared at lower temperatures similar to previously reported SMR behavior. Unlike the dynamical multiferroicity, a large SNMTP response was observed at 5 K as shown in Figure 3 (b-d). It meant that transverse acoustic phonons[14] were the underlying cause of measured response.

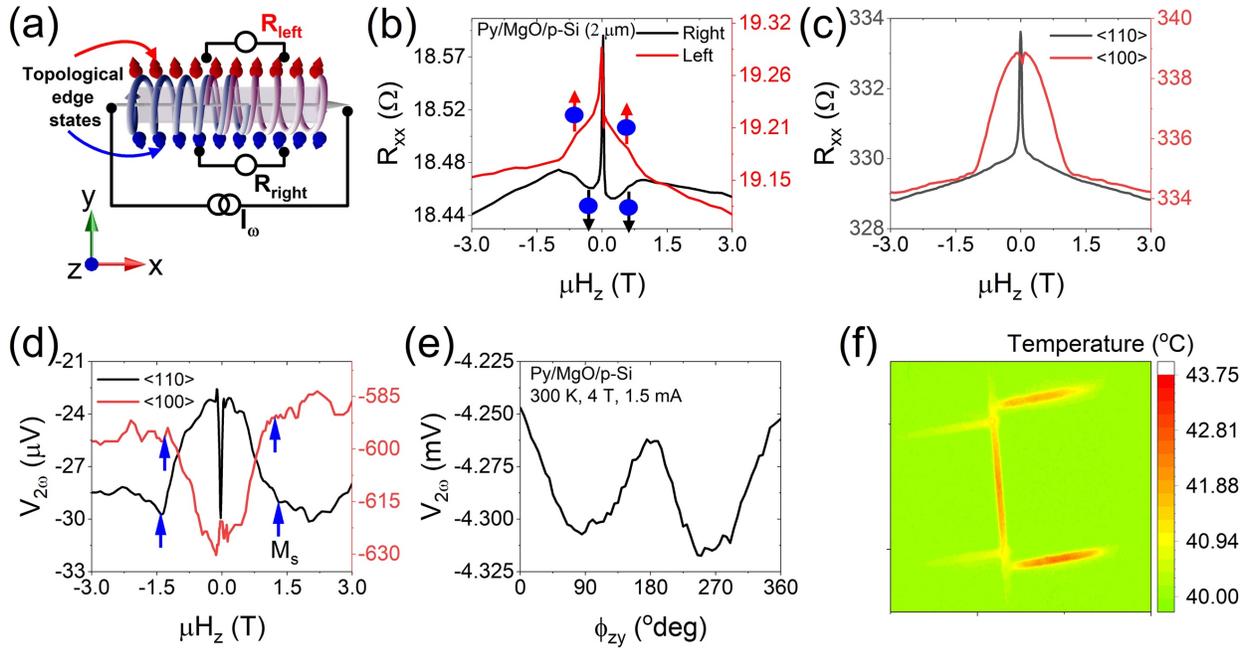

Figure 5. (a) Schematic showing the experimental scheme to measure the resistance on the left and right side of the sample arising due to topological edge states from topological phonons. (b) The longitudinal resistance as a function of out of plane magnetic field from -3 T to 3 T in sample 1 showing the separation of spin-up and spin-down electrons on the right and left side of the sample, respectively. (c) the longitudinal magnetoresistance response in two samples (Py/MgO/p-Si) having longitudinal axis aligned along <110> and <100> crystallographic direction of Si layer, respectively, (d) the longitudinal second harmonic response, attributed to the magneto-thermopower, measured in two Py/MgO/p-Si (2 μm) samples where Si layer was aligned along <110> and <100> crystallographic directions in each. (e) the angle dependent second harmonic response in the Py/MgO/p-Si (2 μm) sample measured at 1.5 mA and 4 T where sample length was 160 μm and (f) the measured temperature profile in the same sample using thermal imaging microscope at 1.5 mA of heating current. The base temperature was 40°C in the thermal imaging microscope measurement.

The spin-momentum locking of the topological phonons should give rise to accumulation of spin-up (+z) and spin-down (-z) electrons on two sides of the sample, again, due to topological edge states from phonon skew scattering and similar to the edge channels in topological insulator[15]. The resulting longitudinal resistance measured on the right and left edge of the sample should have differing contributions from spin-up and spin-down as shown in Figure 5 (a), which would modify the magnetoresistance response. We measured the resistance on left and right edges of the sample 1 as a function of out of plane magnetic field from -3 T to 3 T as shown in Figure 5 (a,b). The magnitude of the resistance was different for left and right measurement, potentially due to asymmetry in phonon skew scattering between spin-up and spin-down electrons. Our assertion was supported by the difference in charge carrier density for positive and negative magnetic fields as shown in Figure 2 (a). The difference in resistance could also be attributed to dimensional imperfections. However, the magnetoresistance response showed different low field behavior (qualitatively) for left and right measurement configuration as shown in Figure 5 (b). This low field behavior was attributed to the accumulation of spin-up and spin-down electrons to right and left side of the sample, respectively, due to scattering from topological phonons in the Si layer as hypothesized and as shown in Figure 5 (a).

The topological properties are a function of material symmetry. Since the phonon dispersion is a function of crystallographic orientation, it meant that inhomogeneity induced topological phonon response should be different along different crystal directions. Hence, we measured the longitudinal magnetoresistance behavior in two samples oriented along <110> and <100> as shown in Figure 5 (c) where strain gradient direction was along <001> in both samples. The two responses were qualitatively different from each other. The difference in response between samples oriented along <110> and <100>, respectively, arose due to dynamical multiferroicity[1,4] since the cross-sectional plane (100) did not contain any <111> temporal magnetic moment direction[4]. It also demonstrated that the low field response, shown in Figure 5 (b), did not arise from Py layer since it was a function of crystallographic orientation of Si layer as shown in Figure 5 (c). We also measured the longitudinal second harmonic response as a function of magnetic field in two samples oriented along <110> and <100> as shown in Figure 5 (d). Both the responses exhibit magneto-thermopower response with change in slope at ~1.25 T, which corresponds to saturation magnetization of Py layer. However, the responses had opposite sign, which also eliminated Py as an underlying cause of the response. Based on the topological edge states and crystallography dependent response, we can conclusively state that the observed behavior arose due to spin-momentum locking of the topological phonons.

The Hall bar samples used in this study had a length of 40 μm between measurement electrodes. In order to eliminate any length dependent behavior, we measured the angle dependent longitudinal second harmonic response in another freestanding sample having length of 160 μm as shown in Figure 5 (e). We also measured the temperature profile using a thermal imaging microscope in the same sample as shown in Figure 5 (f). The thermal imaging microscope clearly showed a parabolic longitudinal temperature distribution in the sample. In spite of that, angle dependent longitudinal response showed a $\sin^2 \phi$ symmetry expected for SNMTP response as shown in Figure 5 (e), which conclusively proved a longitudinal and transverse thermal spin only current.

In conclusion, we have demonstrated spin-momentum locking in the topological phonons in an inhomogeneously strained Si sample. Using location dependent Hall responses, we showed inhomogeneous spatial distribution of the concentration for spin-up and spin-down charge carrier that arises due to spin-momentum locking of topological phonons. We also demonstrated spin-Nernst magneto-thermopower response due to self-heating in a freestanding sample, which would not arise in the absence of spin-momentum locking in topological phonons. The topological phonons also gave rise to edge states similar to topological insulators.

**Acknowledgement**

The fabrication of experimental devices was completed at the Center for Nanoscale Science and Engineering at UC Riverside. Electron microscopy imaging was performed at the Central Facility for Advanced Microscopy and Microanalysis at UC Riverside. Infra-red thermal imaging microscope work was carried out at UC Santa Barbara. SK acknowledges a research gift from Dr. Sandeep Kumar.